\documentclass[12pt]{extarticle}
\usepackage[mathscr]{euscript}
\usepackage{amsmath}
\usepackage[numbers]{natbib}
\usepackage{graphicx}
\usepackage[a4paper, margin=1in]{geometry}
\usepackage{xcolor}

\title{A predictive model for Covid-19 spread applied to eight US states}

\author{Zeina S. Khan,\\
Frank Van Bussel,\\
\&\\
Fazle Hussain$^*$\\
Texas Tech University, Department of Mechanical Engineering\\ 
2703 7th Street, Box: 41021, Lubbock, TX 79409\\
Phone: 832-863-8364\\ 
$*$ fazle.hussain@ttu.edu\\}

\begin{document}
\maketitle
\begin{abstract}
A compartmental epidemic model is proposed to predict the Covid-19 virus spread. It considers: both detected and undetected infected populations, medical quarantine and social sequestration, release from sequestration, plus possible reinfection. The coefficients in the model are evaluated by fitting to empirical data for eight US states: Arizona, California, Florida, Illinois, Louisiana, New Jersey, New York State, and Texas. 
Together these states make up 43\% of the US population; some of these states appear to have handled their initial outbreaks well, while others appear to be emerging hotspots.
The evolution of Covid-19 is fairly similar among the states: variations in contact and recovery rates remain below 5\%; however, not surprisingly, variations are larger in death rate, reinfection rate, stay-at-home effect, and release rate from sequestration. The results reveal that outbreaks may have been well underway in several states before first detected and that California might have seen more than one influx of the pandemic. Our projections based on the current situation indicate that Covid-19 will become endemic, spreading for more than two years. Should states fully relax stay-at-home orders, most states may experience a secondary peak in 2021. If lockdowns had been kept in place, the number of Covid-19 deaths so far could have been significantly lower in most states that opened up. Additionally, our model predicts that decreasing contact rate by 10\%, or increasing testing by approximately 15\%, or doubling lockdown compliance (from the current  $\sim$ 15\% to $\sim$ 30\%)
will eradicate infections in the state of Texas within a year. Extending our fits for all of the US states, we predict about 11 million total infections (including undetected), 8 million cumulative confirmed cases, and 630,000 cumulative deaths by November 1, 2020.
\end{abstract}

\section*{Introduction}
A cluster of pneumonia cases resulting from an unknown pathogen was identified in December 2019 by Chinese Health authorities in the city of Wuhan (Hubei), China \cite{novelcoronavirus}. The pathogen has been identified as a novel strain of coronavirus, SARS-CoV-2 \cite{novelcoronavirus}, now named Covid-19 \cite{world2020coronavirus22}. The exponential growth of the disease prompted Chinese authorities to introduce their strictest level of measures to contain its outbreak, including Wuhan city lockdown, banning public gatherings, shutting down public transport, and prohibiting travel between cities \cite{tian2020early}. Despite these measures, a global pandemic ensued, with 3,855,809  total worldwide cases and 265,861  deaths as of May 9, 2020 as declared by the World Health Organization (WHO) \cite{world2020coronavirus110}. 

The first cases of community transmission in the United States were reported in California, Oregon, Washington state and New York state in late February, 2020 \cite{schu2020how}.  A  Santa Clara, California death on Feb. 6 was deemed the country's first Covid-19 fatality after an autopsy was conducted in April \cite{schu2020how}. A national emergency was declared by US President Donald Trump on March 13, 2020, and testing several days later revealed that Covid-19 had spread to all 50 states \cite{schu2020how}. On March 20, New York City was declared the US outbreak epicenter \cite{schu2020how}. A study, released on April 2020 as a preprint, found via genetic analysis of Covid-19 cases in New York City that the majority of the viruses originated in Europe -- revealing that transmissions had begun as early as January from countries with no travel monitoring \cite{gonzalez2020introductions}. As of June 29, 2020, the US had 2,496,628 confirmed Covid-19 cases, and 125,318 Covid-19 deaths \cite{world2020coronavirus161}. 

Covid-19 challenges faced by the US include fair allocation of adequate medical resources \cite{emanuel2020fair}, minimizing mortality, avoiding overwhelming the health-care system, and keeping the effects of lockdown and social distancing policies on the economy within manageable levels \cite{anderson2020will}. Epidemiological analysis of the virus proliferation is needed to assess the impacts of mitigation strategies including social distancing, sheltering-in-place (voluntary), and quarantines (enforced by authorities) \cite{anderson2020will}, as well as personal habits including frequent hand washing and wearing masks. We have developed a new compartmental  model, extending the  long-standing SIR (Susceptible, Infected, Removed) model \cite{brauer2019simple, small2004plausible, chang2017estimation}, to evaluate and compare several states' responses to Covid-19; with this model we can make estimates, using curve fitting of reported empirical data, of the impact of contact suppression measures and the lifting of such measures. We find that, for the current situation where several states have relaxed stay-at-home measures, Covid-19 will become an endemic virus for at least two years; so it is not surprising that some states are reinstating the stay-at-home measures \cite{lee2020see}. We simulated the effect of fully lifting stay-at-home orders and find that the number of infections increases by approximately an order of magnitude with a secondary peak in several states.  

\section*{A new model}
\begin{figure}
\centering
\includegraphics[width=0.5\linewidth]{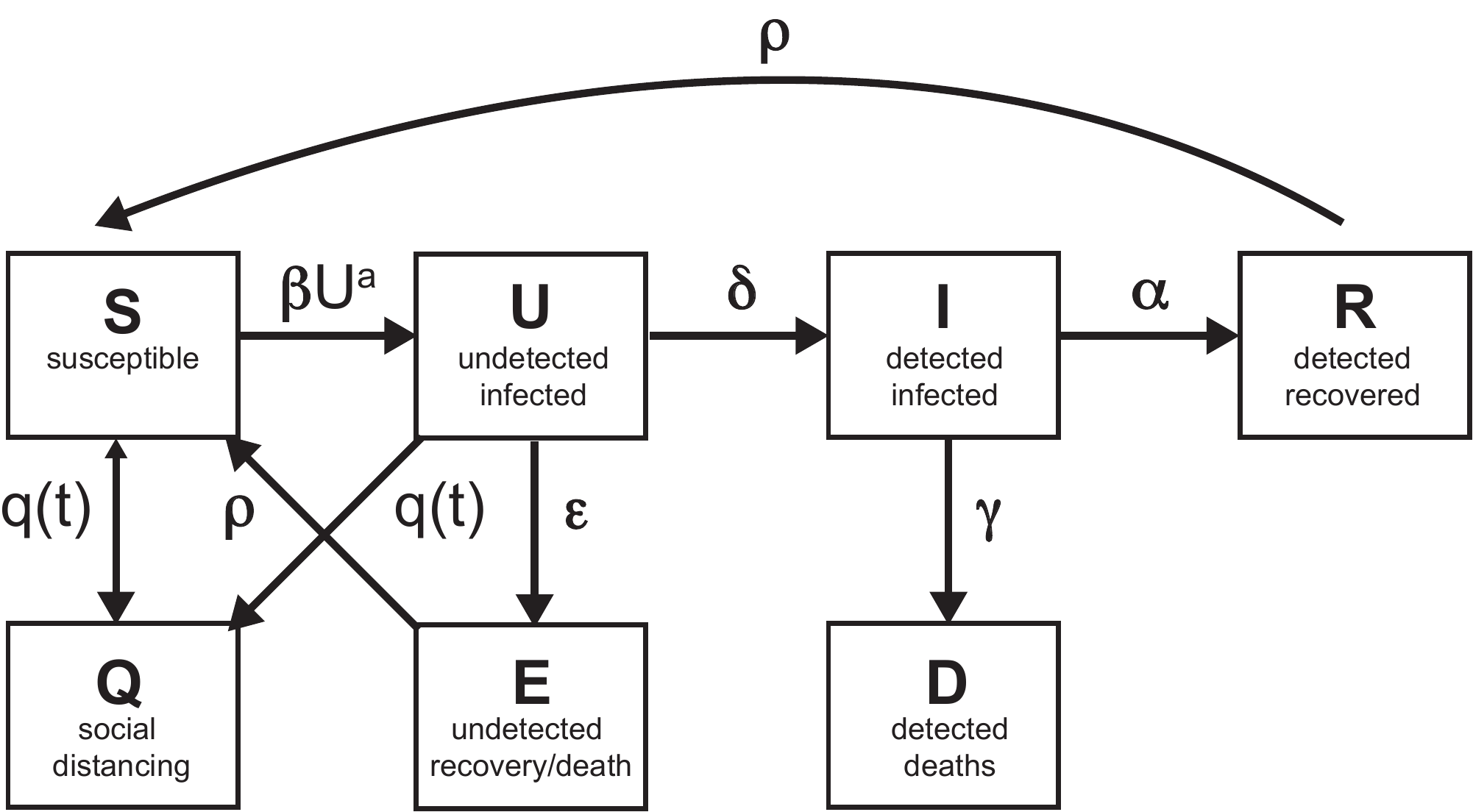}
\caption{Schematic of the compartments, with the rates of transfer between the compartments.}
\label{Ffig1}
\end{figure}
Our new SQUIDER model incorporates additional processes into the classic SIR (Susceptible, Infected, Recovered) model: i) making a distinction between known cases (which are publicly reported) and asymptomatic or mild cases which are not monitored or detected; ii) including the effects of responses, varying by region, to the pandemic, whether direct, through quarantine or medical isolation of diagnosed cases, or less direct such as social distancing efforts; and iii) possible loss of immunity of recovered individuals, allowing some of them to be reintroduced into the Susceptible population. The model thus requires several new compartments, which we will denote as U (Undetected infected), E (Excluded, meaning undetected recovered and possible deaths), and Q (pseudo-Quarantine, a bin to hold a segment of the susceptible and undetected infected populations, that allows us to model lower contact rates due to social distancing). Furthermore, for modeling/fitting purposes we add a separate compartment D for known infecteds who die; deaths from the virus is a statistic that is generally available \cite{Covid19global}. 

The rate equations are as follows:
\begin{equation}
\frac{dS}{dt} = -\beta S U^{a} - q(t) S + \rho(E + R)
\end{equation}

\begin{equation}
\frac{dU}{dt} = \beta S U^{a} - (q(t) + \epsilon + \delta) U
\end{equation} 

\begin{equation}
\frac{dI}{dt} = \delta U - (\gamma + \alpha) I
\end{equation} 

\begin{equation}
\frac{dR}{dt} = \alpha I - \rho R
\end{equation} 

\begin{equation}
\frac{dD}{dt} = \gamma I
\end{equation} 

\begin{equation}
\frac{dQ}{dt} = q(t)(U + S)
\end{equation} 

\begin{equation}
\frac{dE}{dt} =\epsilon U - \rho E
\end{equation} 
Each compartment is normalized by the total population $N$ (includes deaths from Covid-19 but not births and deaths from other causes - see below); hence
\begin{equation}
S+Q+U+I+D+E+R = 1.
\end{equation} 
Note that the coefficients $\alpha$, $\beta$, $\delta$, $\epsilon$, $\gamma$, and $\rho$ are constants (to be evaluated from fits to data).

Before we go through the individual equations we should discuss some of the recurring terms and factors. First, the {\em incidence rate}  $\beta S U^{a}$, the average normalized new infections in time, is nonlinear when $a > 0$. Here $\beta$ is the contact rate, which is the average number of contacts a person has per day, multiplied by the probability of transmitting the disease when contact between a susceptible and an Undetected infected occurs. Detected Infecteds ($I$) are not involved since we assume that, post-diagnosis, the $I$ group are generally in medical isolation or some other form of quarantine \cite{parmet2020Covid}. If $a = 1$, this term describes homogenous mixing of the Susceptible and Undetected infected populations \cite{liu2012infectious}, which may not be accurate for states with isolated populations, low population densities, or many densely populated areas; the relationship may be sublinear or superlinear depending on the population being sparse or dense.  Power law incidence rates (such as $\beta S U^a$) have been shown to improve the accuracy of SIR models \cite{novozhilov2012epidemiological, roy2006representing}.

Second, the factor $q(t)$ models the unavoidable time-dependent social distancing and contact suppression, as well as the releasing of lockdowns, which can be reasonably represented by Gaussian functions. Social distancing is modeled by sequestering a proportion of the Susceptible and Undetetected infected populations at a time $t_1$. This does not imply that some large number of undiagnosed people are put into any actual physical quarantine, only that the available sub-groups for infecting ($U$) and for becoming infected ($S$) are reduced; alternatively, this could be modeled by altering $\beta$ or the power law dependency $a$ in a time-dependent way. Since the model works by transferring populations from one compartment to another over time at different rates, the function $q(t)$ is implemented as a Gaussian pulse centered at a time $t_1$, where $t_1$ is a fit parameter. The standard deviation and height of the pulse are chosen so that a proportion ($q_1$) of the population is sequestered -- this proportion is another fit parameter. This pulse form was chosen, as opposed to a constant value used by some authors \cite{xu2020forecast, moghadas2020projecting, maier2020effective}, because many states went into lockdown on a particular day with numerous people self-isolating \cite{gostin2020presidential, mervosh2020see}. In particular, selecting the day $t_1$ where such a measure is initiated is important as it reflects the compliance of  the state's population. $q(t)$ also has a second, negative, pulse strength modeling the effect of the sequestered population re-entering into the susceptible population. The proportion, $q_2$, of the sequestered people re-entering and the day when this occurs, $t_2$, are also fit parameters.

Equation 1 for the Susceptible population is reduced nonlinearly by new infections $\beta S U^{a}$ due to interactions, and explicitly reduced in a time dependent way by sequestration $q(t) S$ (not by any significant amount until we get close to the activation time $t_1$), as well as increased again by re-entrance at a rate $\rho$ by members of the Excluded ($E$) and Recovered ($R$) groups. This term was added due to the WHO revelation that recovered Covid-19 patients may have little or waning immunity after exposure \cite{world2020immunity}, later confirmed by an antibody study conducted on individuals who had recovered from Covid-19 infections \cite{long2020clinical}. 

Equation 2, for Undetected infecteds, includes the increase due to contact with $S$ members, and removal by various causes. The rate $\epsilon$ closely resembles the recovery rate in the basic SIR model, which (since the model is static,  implying that births and deaths due to other causes are not considered) implicitly includes deaths due to a potentially fatal disease. Note that, at least in the early days of the pandemic, increased overall deaths in comparison with the prior three years were not examined for signs that the virus was active among undiagnosed populations \cite{weinberger2020estimating}. The detection rate $\delta$ specifies the proportion of Undetected infected individuals who are diagnosed with the virus (and are hence no longer undetected); it is added to the $I$ compartment. Finally, this population is also effectively reduced due to social distancing $q(t)$, e.g. residents of many states were encouraged to shelter at home and not seek testing/diagnosis unless they became symptomatic, in order to ease pressure on medical resources \cite{centers2020evaluating}. 

Equation 3 for the detected Infected population, describes increases due to testing at rate $\delta$ and decreases due to death with rate $\gamma$, and recovery with rate $\alpha$. Since these individuals are isolated in designated hospital wards or under quarantine at home \cite{parmet2020Covid}, hence unlikely to be a source of infection to the community at large, we felt there was no need for sequestration (i.e. $q(t)$) when social distancing went into effect. This is in agreement with the WHO guidance on quarantines segregating suspected exposed people \cite{world2020immunity}.

Equation 4 describes the growth of detected Recovered, balanced by outflux of the Recovered population into the susceptible compartment at rate $\rho$ due to little or waning immunity, expected for human coronaviruses \cite{callow1990time}. Equation 5 describes the increase in Deceased detected individuals. Equation 6 describes the increase in the pseudo-Quarantine compartment due to official contact suppression measures $q(t)$, which as stated above is only significant around the activation time $t_1$. Of course, one expects that the coefficients for $U$ and $S$ could be different, but are kept the same ($q(t)$) for simplicity, discussed later. Equation 7 describes increases in the undetected recovered population at rate $\epsilon$, and decreases in this population due to loss of immunity of $E$ at rate $\rho$. Equation 8 is the trivial statement that the sum of all of these compartments ($S$, $Q$, $U$, $I$, $D$, $E$, $R$) add up to the total population $N$.

\section*{Assumptions}
Several simplifying assumptions or idealizations have been made. To begin with, in our model the detected Infecteds ($I$) do not transmit the disease to the Susceptible ($S$) population. It is generally the case that in all such disease outbreaks (e.g. the 2014--2016 Ebola outbreak), even when strict quarantine measures are in place, medical service providers and other people rendering direct aid to victims are themselves vulnerable to infection; when the outbreak (in the non-healthcare worker population) is contained they may even make up the substantial proportion of cases \cite{centers2019ebola}. However, in the current situation where the disease circulates through the general population, and safety protocols (such as restricting visitation by friends and relatives) are rigorously enforced by health providers, the number of such cases is statistically negligible.

Other simplifying assumptions: as mentioned above, we have opted to keep our contact rate $\beta$ constant and instead vary the $S$ and $U$ compartment population levels to mimic social distancing effects (for example staying at home). In future versions of our model we may incorporate time dependent $\beta$ or $a$ in order to disentangle population wide transmission suppression (e.g. face masks and other protective gear for the public) from social contact suppression (cancellation of concerts and other public gatherings), but for the sake of simplicity in both coding and analysis we have opted for now to use only one time dependent rate. In the same vein, the $q(t)$ function removes Susceptibles and Undetetected infecteds at the same rate (we have no reason at this time to differentiate the rates of change of these populations). Similarly, people from the Excluded $E$ and Recovered $R$ compartments lose immunity at the same rate $\rho$. Indeed, it is possible that people who experienced milder forms of the disease ($E$) lose immunity faster than those who experienced more severe symptoms and sought out treatment ($R$) \cite{long2020clinical}; however, we use a single rate 
$\rho$ for simplicity. Finally, we do not consider the effects of births, vertical transmissions, immigrants, emigrants, or deaths due to other diseases or trauma. The inclusion of deaths due to diagnosed virus cases makes the model not entirely static, but the disease's total deaths as a proportion of the total population is low enough that births and other such aspects can be safely omitted.

\section*{Methods}
All numerical simulation for equations 1-8, fits, and data management were done in {\sc Matlab}.  Data for cumulative confirmed cases and deaths were obtained from the Johns Hopkins University (JHU) Center for Systems Science and Engineering, which has been making highly credible US and global Covid-19 time-series statistics available to the public on the  GitHub \cite{csse2020data} website. Raw data in the original CSV files were converted to Matlab {\tt table} data structures for ease of access; since the US data were broken out by municipality/county, it was necessary to aggregate this to create each state wide time series. 2019 estimates of state populations used for normalization were acquired from the US Census Bureau \cite{uscb2020data}.

Least squares fits of numerically generated curves to the data were obtained with Matlab's {\tt lsqcurvefit} using a {\em trust-region-reflective} algorithm \cite{ml2017lsq}. One of the benefits of this fitting routine is that fit parameters can be given bounds or fixed values (the latter being especially useful during model development and testing). Fit parameters were: all model rate parameters ($\beta$, $\epsilon$, $\delta$, $\alpha$, $\gamma$, and $\rho$), power law exponent $a$, initial condition for unknown infecteds $U(0)$, plus two pairs of parameters governing sequestration of populations due to social distancing -- peak $q_i$ values and dates of application $t_i$ where $i = 1,2$. Fits were done in two stages. When work was begun on this project, we had data from Jan 22 to May 9, which allowed us to make initial fits for all rate parameters, infectious power, initial conditions, and lockdown effects. In the course of writing the results up, we realized we would need to incorporate the effect of breaking developments (cessation of state stay-at-home orders and new spikes in cases), so a second fit  was done to determine the time and magnitude of the release of lockdown ($q_2$, $t_2$) keeping all of the previous parameter values unchanged.

Since all the rates are bound between 0 and 1, the $t_i$ parameters were rescaled by the total time of the simulation to fall within the same range (this helps the fitting routine when determining step sizes while revising the current solution). Fits were made comparing certain selected and/or aggregated simulation results against normalized JHU data for cumulative confirmed cases and deaths, simultaneously. {\tt lsqcurvefit} default values were used for tolerances, iterations, and step size, but because the proportions of state populations were so small simulation results and normalized data were rescaled to increase the magnitude of the error, preventing the fit routine from prematurely settling for a solution (the scaling formula was chosen so that the norm of the rescaled test data was equal to the number of elements in the test data matrix).

Simulation results were generated using Matlab's {\tt ode45} function, which uses an explicit Runge-Kutta $(4,5)$ formula. Default settings were used for the solver, except that the maximum step-size was constrained to be $\leq 0.5$ days (this prevents the solver, which uses an adaptive step-size, from accidentally stepping over the sequestration date $t_1$). The implementation of the model itself, coded in a function that is given as an argument to {\tt ode45}, is for the most part straightforward; the only aspect that requires any further comment is the handling of the sequestration function $q(t)$.

As mentioned above, the first sequestration is modeled as a smoothly shaped pulse centered at time $t_1$. Since the system of ODEs works by transferring populations at various rates between compartments, in practice we set a temporary rate which changes every time the solver calls the model subroutine. For this we use the value of a Gaussian curve at $x=t$ with mean value $t_1$, height normalized, and width set so as to achieve the sequestration we desire (easily calculated using Matlab's {\tt normpdf} function). Since {\tt ode45}'s adaptive step-size decreases when it detects unexpected movement in the $q$ rate, a well resolved sampling of different values near the peak $t_1$ is obtained. To make analysis more straightforward, we decided that the peak rate given to the solver would be equal to the total sequestration effect (e.g., if we set the peak rate to 0.15, then roughly 15\% of the $S$ and $U$ compartments would ultimately be moved into the $Q$ compartment). To achieve this, for peak rates $\leq 0.625$ we normalized the height to the peak rate, and set the standard deviation to a value between 0.4 and 0.625 determined by trial and error and fit to a cubic polynomial. For peak rates above 0.625, a much more complicated formula was needed to achieve the desired effect; since such high sequestration never appeared in any of the fits we will omit any further discussion of this, except to say that to get effective clearance (99.84\%) of the entire relevant compartments we use a Gaussian with both height and width set to $\approx 1.6$.

Lastly, for the purpose of doing projections the model implementation subroutine accepts an arbitrary number of peak rate/activation date pairs tacked onto the end of the parameter vector it takes as one of its arguments. This allows us to test the effect of doing several interventions of possibly different magnitudes. Also, a negative peak rate is implemented as returning the specified proportion of the sequestered population in $Q$ back to $S$ (since {\tt ode45} has no facility for keeping track of the ratio of $S$ and $U$ populations that were originally sequestered, it was felt that returning everyone to $S$ was the most sensible approach). To make this feasible codewise, at any particular time $t$ only the $q_i$ with peak time $t_i^*$ closest to $t$ is executed. In practice, if peaks are set too close to each other (e.g. within half a week or so) they may interfere with each other's ability to achieve full sequestration or release; but since this is essentially the case in real life as well we thought it not to be a priority to address this issue.
\section*{Results}

\subsubsection*{Coefficient evaluation}
\begin{figure}
\centering
\includegraphics[width=0.75\linewidth]{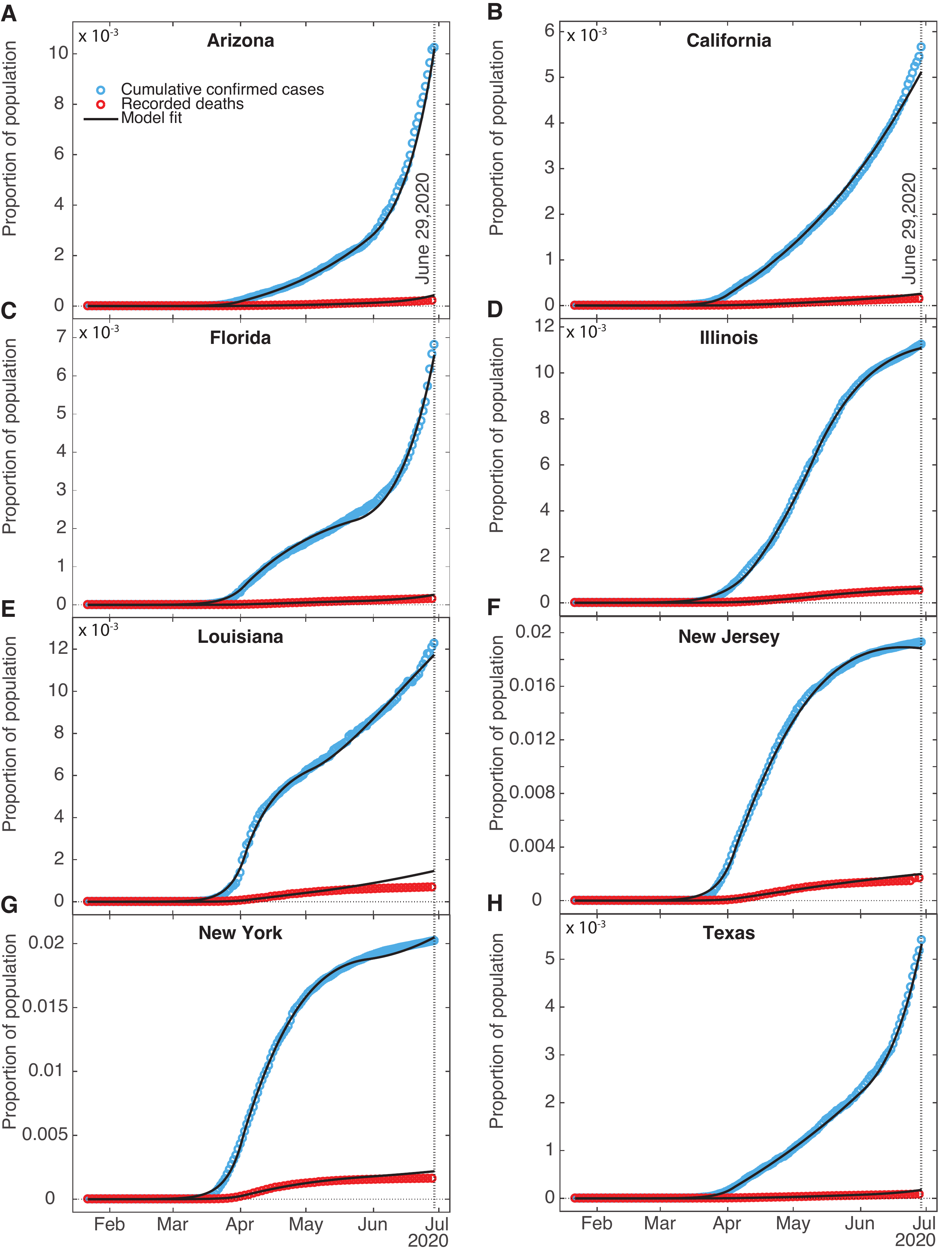}
\caption{{\bf SQUIDER model fits.}
Fits of our compartment model to recorded data on confirmed cumulative case counts and deaths for (A) Arizona, (B) California, (C) Florida, (D) Illinois, (E) Louisiana, (F) New Jersey, (G) New York state, and (H) Texas. Fits to all states' confirmed case counts and deaths have $R^2$ $\geq$ 0.996, implying extremely good fits. Data was obtained from The Johns Hopkins University \cite{csse2020data}. The vertical dashed line indicates the last date fitting data was obtained for.}
\label{Ffig2}
\end{figure}

Figure \ref{Ffig2} shows fits of the model to data (cumulative confirmed case counts and deaths due to Covid-19) for Arizona, California, Florida, Illinois, Louisiana, New Jersey, New York State, and Texas. The fits to cumulative case counts are all excellent -- even for states such as Illinois, New Jersey, and New York that have a distinct inflection in case counts. It is also apparent that some states, such as Arizona, Florida, Louisiana, and Texas, have had rapidly rising case counts in June -- likely due to easing of restrictions. This feature is captured by our $q(t)$ -- where application of a second, possibly negative pulse moves people from the Q to the S compartments as described above. Note that model fits to cumulative death counts deviate from the data in some states in June (such as Louisiana, New York, and Texas). This may be due to the postulated lack of reliability in confirmation of US Covid-19 deaths, which may be  significantly undercounted \cite{bump2020fau}. Fit parameters are listed in Table 1. The contact, detected recovery and exclusion rates vary by less than 5\%; however, reintroduction rate varies by 320\%, re-release rate by 268\%, stay-at-home effect by 103\%, and death rate by 59\%. These variations, though large, are not surprising due to different states having been in different stages of the outbreak. 

We compare our fit parameters with the classical SIR model which, to remind the reader, involves only Susceptible, Infected, and Recovered compartments. The $\beta$ and $\epsilon$ parameters correspond to the contact and recovery rates of the SIR model; the fit values imply that for unconstrained epidemic situations (with $q(t) = \delta = 0$) the disease has a reproduction number $\mathscr{R}_0 = \beta/\epsilon$ of around 5 and a duration of about a week, consistent with prior investigations \cite{maier2020effective,liu2020reproductive}.

It is surprising that the detection rate $\delta$ is so high for all of the states ($\approx 0.5$); however, a recent nationwide coronavirus antibody study by the Spanish Health Ministry \cite{jones2020spanish} suggests that the number of unknown infected and unknown recovered in large and heterogeneous jurisdictions, while significant, is not orders of magnitude larger than the number of confirmed cases. This goes against some prior speculation that the asymptomatic and undiagnosed cases might be as much as 10 times the official count \cite{gaeta2020simple}, which would suggest a detection rate 5 times smaller than our $\delta$ values. 

Returning back to figure \ref{Ffig2} and table \ref{Ttab1}, the death rate from diagnosed cases $\gamma$ falls between around 2\% and $3\frac{1}{2}$\%, which is well within the quite wide range of case-fatality rates reported for earlier phases of the pandemic \cite{owid2020mortality}. The recovery rate $\alpha$ for diagnosed cases, seems somewhat high (in the 0.45 -- 0.5 range for New York and New Jersey, and around 0.7 -- 0.8 for the other states); this could reflect the fact that detection of any disease would normally occur after that disease has already partly run its course, but it should be kept in mind as well that this compartment has a minimal effect on the size of the fitted compartments $I$ and $D$, so the fitting routine may not be as constrained in selecting the $\alpha$ value as the other fit parameters. The re-entry (due to loss of immunity) rate $\rho$ is fairly low for most of the states, which indicates that this is not a significant factor for the initiation of the outbreak. Such low $\rho$ values are not unreasonable since loss of immunity to corona viruses that cause common colds is typically slow, even taking months \cite{callow1990time}. Indeed, one recent report found that antibodies in a high proportion of individuals who recovered from Covid-19 started to decrease within 2--3 months after infection, consistent with our findings \cite{long2020clinical}. As we see below, this slow reintroduction of people to the Susceptible compartment obviously results in an endemic infection in the predictions.

Our fitted peak initial sequestration values range over 0.1 -- 0.2 for all states except Illinois (whose value is close to 0.04); it should be kept in mind this is not the actual quarantining of approximately 15\% of the population of most states, but rather the effect of overall reduction in social contacts between susceptible and infectious populations, which in the US was never pursued as systematically or consistently as in some other jurisdictions. The parameter $t_1$ enables prediction of what day self-isolation policies started having significant effects on case counts -- see Table 1. To compare $t_1$ to states' directives, stay-at-home orders were issued in Arizona on March 31, California on March 19, Florida on April 1, Illinois on March 21, Louisiana on March 23, New Jersey on March 21, New York on March 22, but was suggested in Texas on April 2 \cite{merv2020reopen}. The one week or so time lag between action and its effect  should be expected to be related to the infection lasting about a week as mentioned above. It is possible that Texas, Arizona, and Florida residents were following local orders which prescribed sheltering in place sooner than the  state orders: as examples Dallas had a state of emergency declared on March 19, and Houston residents were urged to stay at home on March 24 \cite{deb2020harris}.

Many states partially re-opened in May. Our fit parameter $q_2$ for the various states corresponds to the percentage of people moving between the Susceptible and Quarantine compartments -- where a negative number indicates the percentage decreasing, moving from the Quarantine to the Susceptible compartment. Specifically, Arizona, Florida, and Texas had significant numbers ($>$30\%) exiting from stay-at-home conditions, whereas California and Louisiana had smaller numbers ($<$11\%), Illinois had 4$\frac{1}{2}$\% additional people enter the Q compartment, and New Jersey had a small Q entry ($<\frac{1}{2} $\%). The date at which this occurred, $t_2 $, can also be compared with the dates stay-at-home orders were lifted or expired -- also in Table 1. Stay-at-home orders were relaxed or expired on May 15 in Arizona, low risk businesses and some restaurants opened in California on May 8; stay-at-home expired in Florida on May 4, in Louisiana on May 15, in New York on May 28, and in Texas on April 30. The time lag between $t_2$ and official dates may correspond to the end of the school year and people's perception of safety.

\begin{table}
\begin{center}
\resizebox{\linewidth}{!}{%
\begin{tabular}{ |c|c|c|c|c|c|c|c|c| } 
 \hline
 parameter & Arizona & California & Florida & Illinois & Louisiana & New Jersey & New York & Texas \\ 
  \hline
  \hline
 $\beta$ & 0.7497 & 0.7482 & 0.7439 & 0.7329 & 0.7589 & 0.7614 & 0.7644 & 0.7485 \\ 
  \hline
 $a$ & 0.9942 & 0.9943 & 0.9944 & 0.9841 & 0.9948 & 0.9966 & 0.9970 & 0.9943 \\
  \hline
 $\epsilon$ & 0.1548 & 0.1551 & 0.1610 & 0.1916 & 0.1441 & 0.1390 & 0.1361 & 0.1551 \\
  \hline
 $\delta$ & 0.5038 & 0.5035 & 0.5038 & 0.5774 & 0.5009 & 0.5073 & 0.5067 & 0.5036 \\
  \hline
 $\alpha$ & 0.7086 & 0.7324 & 0.7910 & 0.8696 & 0.7115 & 0.5104 & 0.4426 & 0.7404  \\
  \hline
 $\gamma$ &  0.0297 &  0.0249 &  0.0229 &  0.0346 &  0.0213 &  0.0243 &  0.0339 &  0.0191 \\
  \hline
 $\rho$ & 0.0067 & 0.0162 & 0.0176 & 0.0094 & 0.0750 & 0.0230 & 0.0084 & 0.0144\\
  \hline
 $q_1$ & 0.1384 & 0.1384 & 0.1547 & 0.0428 & 0.1728 & 0.1561 & 0.1894 & 0.1455 \\
  \hline
 $t_1$ (2020) & Apr. 01 & Apr. 02 & Apr. 04 & Apr. 15 & Apr. 04 & Apr. 05 & Apr. 02 & Apr. 04 \\
  \hline
 $q_2$ & -0.5000 & -0.0087 & -0.5439 & 0.0451 & -0.1051 & 0.0036 & -0.2630 & -0.3807 \\
  \hline
 $t_2$ (2020) & May 30 & May 30 & May 24 & May 12 & May 04 & Apr. 10 & May 29 & Jun. 03 \\
  \hline
 $U(0)\times N$ & 0.0055 & 0.0511 & 0.1416 & 0.0957 & 0.0263 & 0.2099 & 0.6826 & 0.0212 \\
 \hline
\end{tabular}}
\end{center}
\caption{SQUIDER model fit parameters for selected US states. $t_1$ and $t_2$ values are days counting from January 22, 2020.}
\label{Ttab1}
\end{table}
The initial values of the undetected infecteds $U(0)$ are all less than one individual (some significantly), implying that none of the states we look at had any actual cases on January 22 (the first day for which we have data). While the ODE results can be scanned to find an estimate for the arrival of the first case (or first two, or five cases) in a jurisdiction, some caution should be exercised in applying this number, since magnitudes at this point are still too small to make statistically valid comparisons. Given that the model estimates there to have been at least 10 cases in the states studied (except Arizona, Florida, and Illinois) by the 3rd or 4th week of February (the 1st week of February for New York) we think it is probable that Covid-19 was spreading considerably sooner in New York, New Jersey, Texas, and Louisiana than previously assumed. This  implies that stronger measures -- such as travel bans, cluster identification, contact tracing, and quarantine measures -- were needed to fully contain the outbreak \cite{pinotti2020lessons, chinazzi2020effect}.  California had reported cases already in late January, yet both the reported data and our ODE model show the main outbreak occurred well after New York or New Jersey. This is likely if  the western states were dealing successfully with cases coming directly from Asia, but lost control of the outbreak when infected individuals started arriving from the eastern US or possibly Europe; possible differences between the Asian and European Covid-19 strains are are not addressed here.
\begin{figure}[htb!]
  \centering 
 \includegraphics[width=0.75\linewidth]{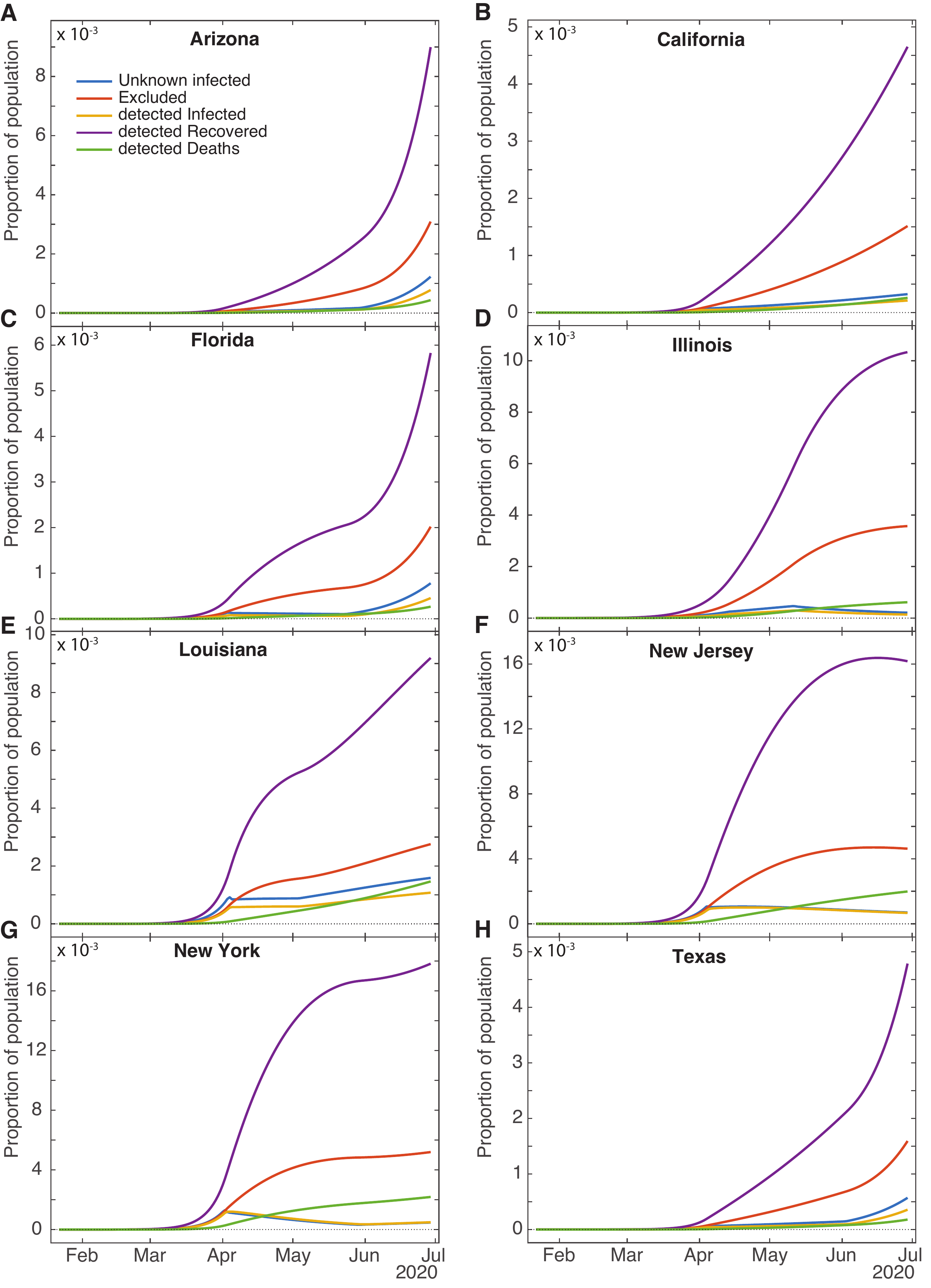}
  \caption{{\bf Computed results for the other compartments for different states.}
Unknown infected (U), Excluded due to undetected recovery and deaths (E), detected Infected (I), detected Recovered (R), and detected Deaths (D). 
 }
\label{Ffig3}
\end{figure}

  Figure \ref{Ffig3} shows the behaviors of the $U$, $E$, and $R$, as well as $I$ and $D$ (fitted compartments) over the fit period (January 22 to June 29, 2020). Note that the most striking difference in the shapes of the curves among the states is that between Arizona and New Jersey -- the other states lie in the range between them. It appears that sequestering $S$ and $U$ populations had a much stronger effect in New Jersey than in Arizona on Recovered and Excluded populations -- with plateaus in these populations becoming apparent in May 2020. Notably, the state of New Jersey has not yet announced plans to reopen. Similarly, the Unknown infected and confirmed Infected cases rise much less quickly in and beyond April, while the confirmed Deaths continue to rise due to active infections. In contrast, in Arizona, the values for all of the compartments are approximately 50\% smaller, though all of them increase with a steady apparent slope in June, suggesting that the outbreak is at an earlier stage and is likely fueled by increasing contacts due to easing stay-at-home measures \cite{lee2020see}. The effect of applying contact suppression measures (pseudo-quarantine) becomes apparent on day $t_1$ in $U$ and $I$ curves in most of the states as it forms a small hump in the curves (see figures \ref{Ffig3}C-G). Notably, the states whose undetected infected population $U$ decreases after quarantine and its release are Illinois, New Jersey and New York -- these states put in strict shelter at home measures \cite{tog2020bay}, whereas,  as an example, the Texas governor has not yet made staying at home mandatory \cite{svit2020gov} and permitted its residents to attend in-person church services \cite{link2020fact}.  The rapid rise in $U$ and $I$ in June in Arizona, Florida, Louisiana, and Texas is likely due to early opening up the states' economies, such as allowing in-person dining and retail shopping as examples \cite{lee2020see} -- reflected in their higher magnitude $q_2$ values.

\begin{figure}[htb!]
  \centering 
 \includegraphics[width=0.65\linewidth]{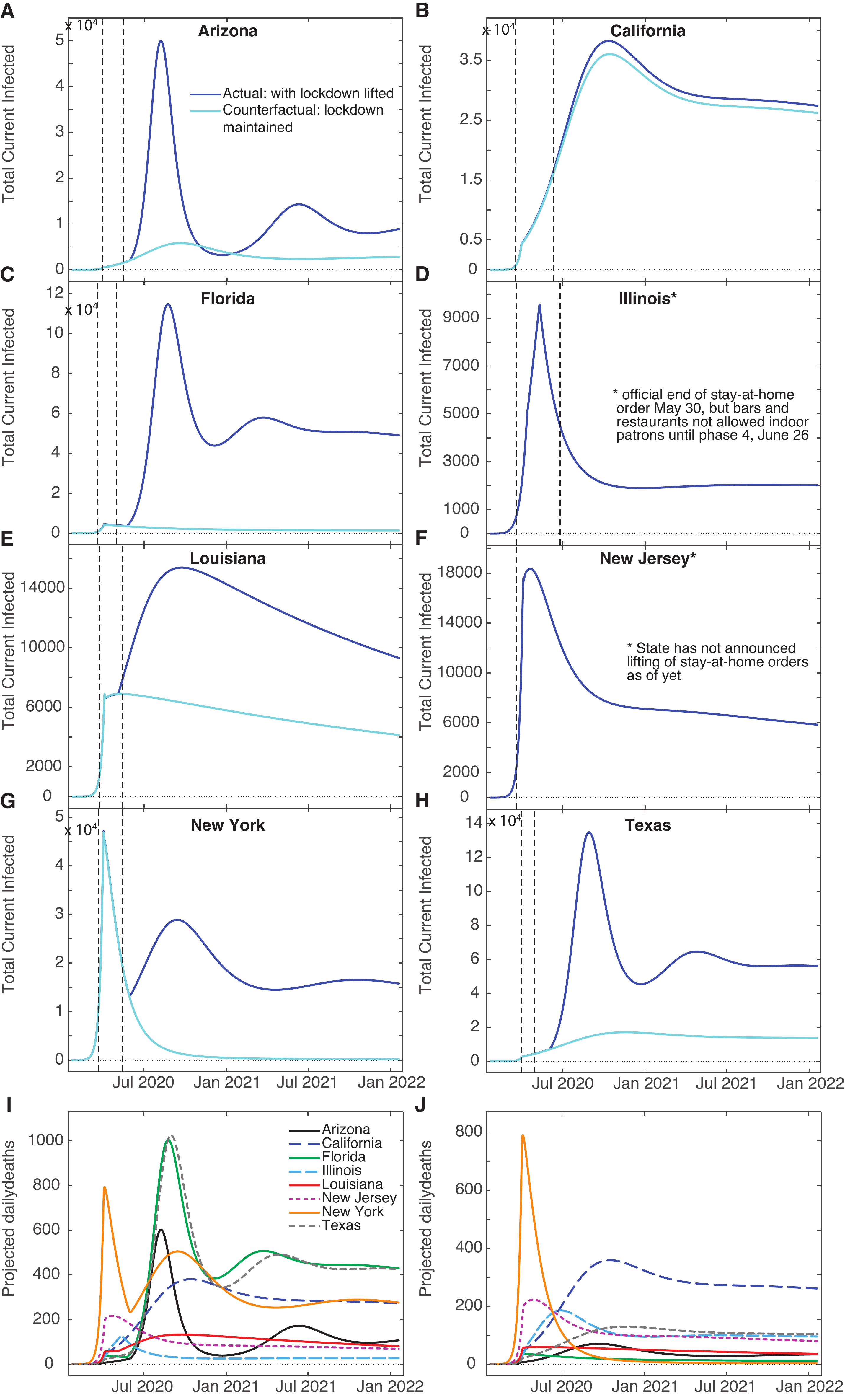}
  \caption{{\bf Model predictions.}
Total infected (U+I), (I), and confirmed daily deaths (D) for two years beyond the first day of recorded infections generated from our model fits, and the counterfactual scenario of not having lifted stay-at-home orders for: (A) Arizona, (B) California, (C) Florida, (D) Illinois, (E) Louisiana, (F) New Jersey, (G) New York, and (H) Texas. The left vertical dashed line indicates the day "shelter-in-place" orders were implemented, and the right vertical dashed line marks the day where such orders were lifted. (I) Confirmed daily death counts for each state for model fits. (J) Confirmed daily death counts for the counterfactual scenario of not lifting orders.
 }
\label{Ffig4}
\end{figure}

\subsubsection*{Model Predictions}
We have generated future predictions using the current  fits, out to two years beyond the first date of recorded US cases. Figures \ref{Ffig4}A-H show that the total number of infections increases substantially in most states from July 2020 to January 2021, except for Illinois, which apparently experienced its peak case count in May. It is predicted, however, that all states will have continued Covid-19 infections for the next two years, some with small secondary peaks occurring in the spring of 2021. California, Illinois, Louisiana, and New Jersey do not have secondary peaks -- this is likely due to these states having negative $q_2$ values (meaning more people enter $Q$) close to the reintroduction rates $\rho$, or very small $\rho$ values. For oscillations to be present in compartment models, cycling of populations has to occur at an intermediate rate -- having a high re-entry rate leads to steady infections; and having no re-entry results in eradication of the virus. Our projected daily deaths (figure \ref{Ffig4}I) show that Arizona,  Florida, New York, and Texas have secondary peaks in deaths after the first main peak. New Jersey and Illinois avoid a secondary peak, presumably because these states haven't yet reopened. We have performed fits to all of the states in the US, and predict 11,326,089 cumulative cases, 8,346,433 cumulative confirmed cases, and 633,562 cumulative deaths by November 1, 2020 (for all fits $R^2 \geq 0.96$).

We have also generated counterfactual estimates of case counts (figures \ref{Ffig4}A-H) for the hypothetical situation where stay-at-home orders were not lifted. The daily peak total cases is $\approx$ 10 times lower for Arizona, Florida, Louisiana, New York, and Texas, in comparison with trends predicted from current data. Keeping the stay-at-home orders had weaker effects in Louisiana and California than the other states -- due to only releasing small numbers of their $Q$ population back to $S$ ($\approx$ 10\% for Lousiana, $<$1\% for California). Our counterfactual daily deaths (figure \ref{Ffig4}J) show that maintaining staying-at-home results in significantly reduced deaths in Arizona, Florida, Louisiana, New York, and Texas. California was not strongly affected because its residents did not fully re-open, and Illinois and New Jersey's counterfactual and factual projections do not differ since these states did not reopen.

\begin{figure}[htb!]
  \centering 
 \includegraphics[width=0.8\linewidth]{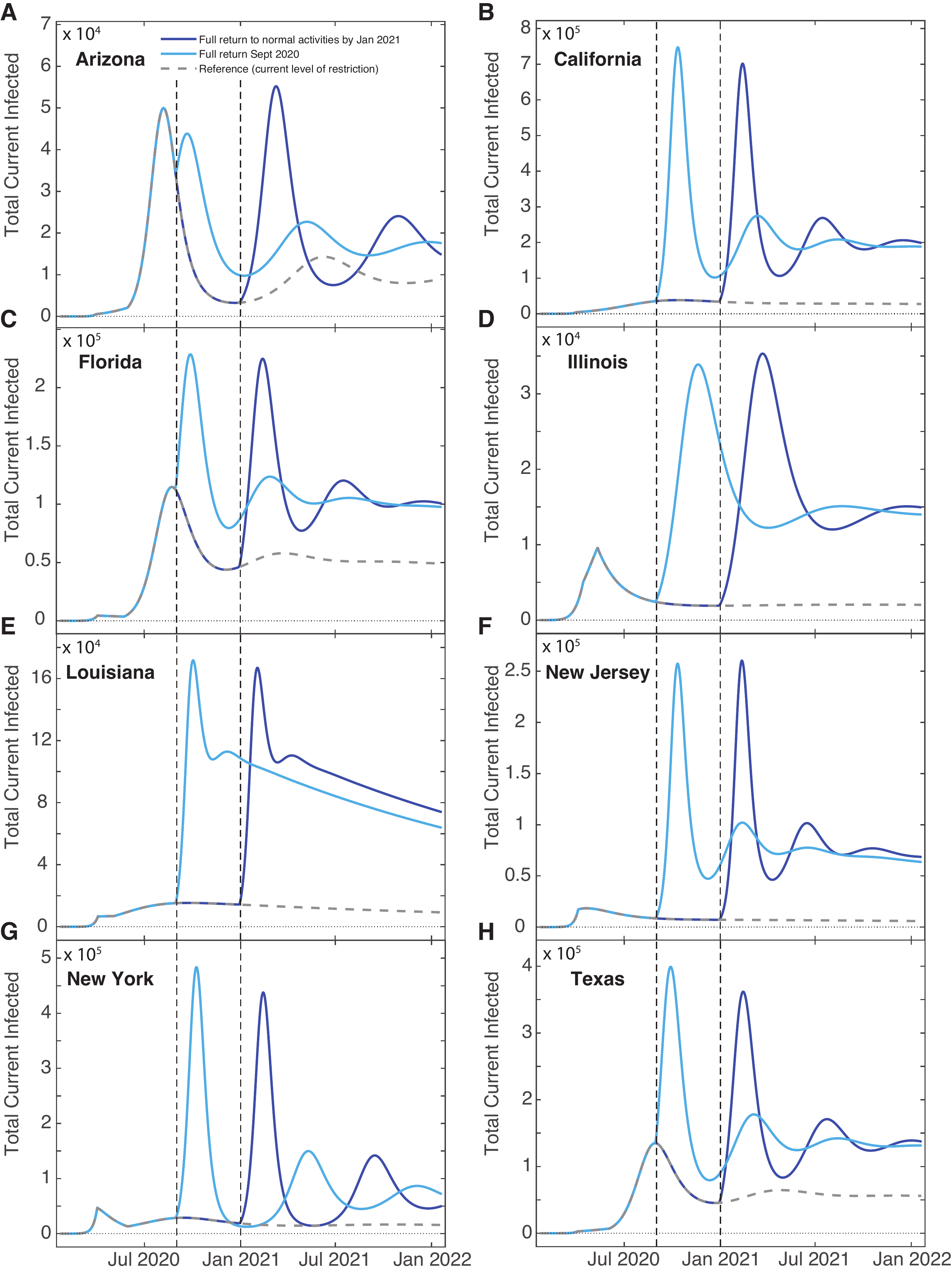}
  \caption{{\bf Predictions for full state re-openings.}
Total infected (U+I), for full state re-openings on September 1, 2020 (light blue) and January 1, 2021 (dark blue), shown alongside our current forecast (gray dashed line) for: (A) Arizona, (B) California, (C) Florida, (D) Illinois, (E) Louisiana, (F) New Jersey, (G) New York, and (H) Texas. Vertical dashed lines delineate the day of lockdown release.
 }
\label{Ffig5}
\end{figure}
\subsubsection*{Effects of full re-opening}
We further considered the effect of fully re-opening the states selected by transferring all of the $Q$ compartment populations into the $S$ compartment on two future dates -- September 1, 2020 and January 1, 2021 -- see figure \ref{Ffig5}. These dates were arbitrarily selected, but demonstrate an apparently universal behavior. We observe that, immediately after the releases, the number of total daily infections rapidly rise to a peak, and secondary peaks afterwards are also observed. Surprisingly, some states such as Arizona and Illinois experience a larger peak with reopening later -- we have no explanation for this surprising behavior. For most states, including California, Florida, Louisiana, New York, and Texas, reopening later slightly decreases peak case counts. Reopening later had little effect on New Jersey's peak daily total cases.

\begin{figure}[htb!]
  \centering 
 \includegraphics[width=0.45\linewidth]{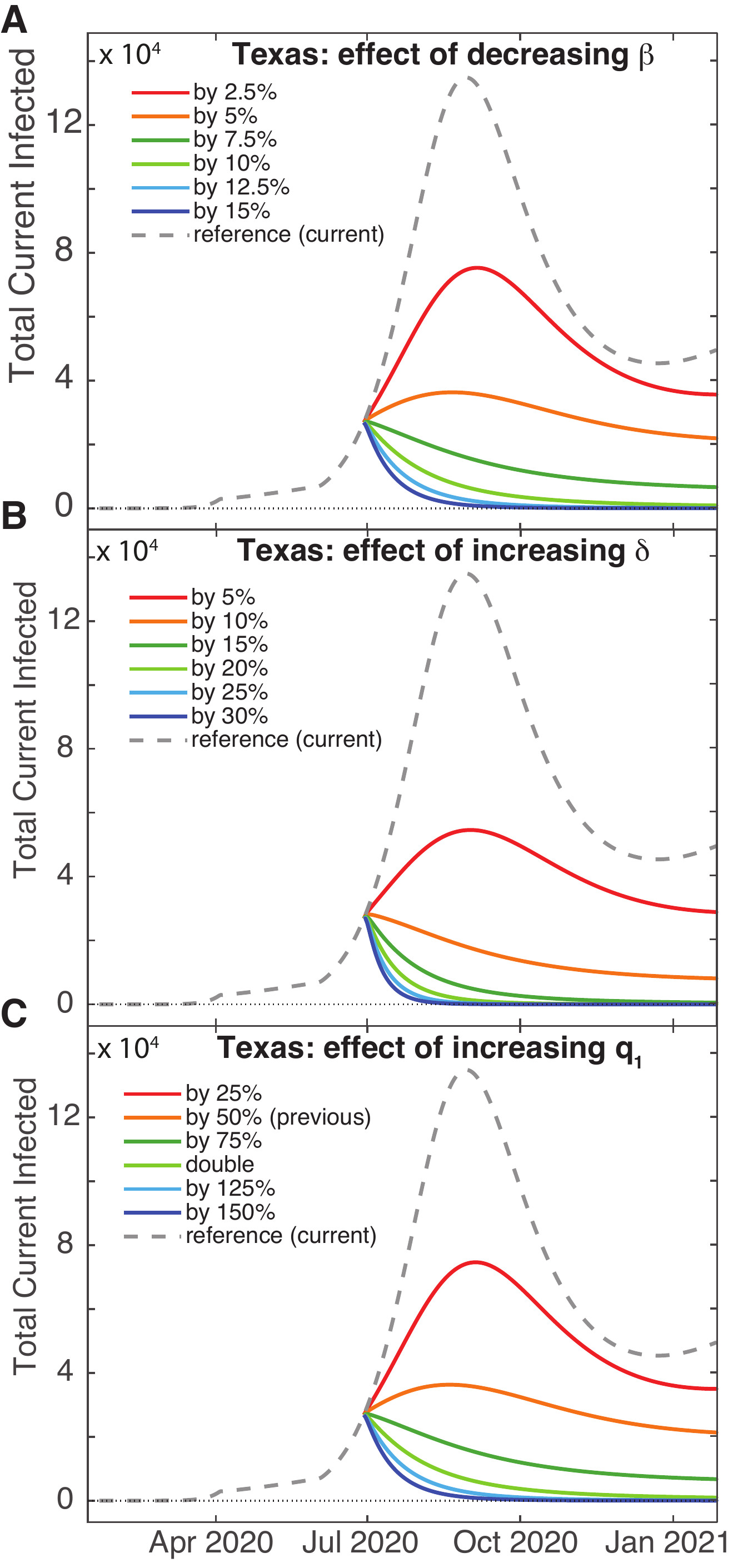}
  \caption{ {\bf Predictions for increasing the effect of non-pharmaceutical interventions for Texas.} Total (U+I) case counts for: (A) Decreasing the contact rate $\beta$, (B) increasing the detection date $\delta$, and (C) increasing the sequestration compliance via quarantine rate $q_1$ .
 }
\label{Ffig6}
\end{figure}
\subsubsection*{Sensitivity to intervention level}
Given our grim predictions (in figures \ref{Ffig4}-\ref{Ffig5}), it is natural to ask if there is some non-pharmaceutical intervention (i.e. vaccines) that could improve the situation. We show the effects of increased social distancing and mask-wearing by reducing the contact rate $\beta$ in figure \ref{Ffig6}A for the state of Texas. Decreasing $\beta$ by 10\% results in virtual eradication of Covid-19 in Texas within one year. Increasing detection rate (i.e. testing) by approximately 15\% will also eradicate the virus within a year, shown in figure \ref{Ffig6}B, as will increasing  lockdown compliance $q_1$ by 100\%. These show that the most sensitive intervention to reduce this virus is to reduce contact rates.

\subsection*{Discussion and Conclusions}
Several studies in the peer-reviewed literature and various pre-print servers have modeled the growth of Covid-19 infections and deaths in the US or its various states. One of these uses a SEIR model (Susceptible, Exposed, Infected, Removed) implemented on a network to simulate inter-state travel \cite{peirlinck2020outbreak}. They predict that, in the absence of countermeasures, the outbreak peaks on day 54 in their simulation (May 10, 2020). They also predict that on day 50, the number of cases has decreased in New York, New Jersey, Washington, and Louisiana whereas Texas and California have many increasing cases. Other SEIR-type models with additional compartments \cite{xu2020forecast, moghadas2020projecting} including quarantine, also predict that the US outbreak peaks near May 10, 2020 \cite{xu2020forecast}, or peaks in the general population 15 weeks into the outbreak (approximately by the last week of April) if only 5\% of the population practices self-isolation within a day of symptom onset; if 10\% of the population self isolates this pushes the peak by an additional 3 weeks, demonstrating that small variations can have significant effects \cite{moghadas2020projecting}. 

A Covid-19 SEIRS model (where recovered become susceptible again) including co-infection with additional human coronavirus strains and a periodic basic reproduction number $\mathscr{R}_0$ corresponding to seasonal forcing, combined with US data, predicts that wintertime outbreaks will occur for several years if immunity wanes -- as also occurs with other coronaviruses \cite{kissler2020projecting}. This study also predicts that the number of confirmed Covid-19 cases in the first wave strongly depends on the peak value of $\mathscr{R}_0$. Furthermore, social distancing was  tested by reducing $\mathscr{R}_0$; applied once this may push the epidemic peak to the autumn, whereas intermittent application can reduce the total number of cases \cite{kissler2020projecting}. 

As expected for nonlinear systems, some predictions, even if accurate for a short time, can deviate significantly with increased time. A logistic model of Covid-19 growth in the US predicts that the cumulative number of cases plateaus by May 14, 2020 \cite{kriston2020projection}. Alternatively, a neural network parametric model was developed, which predicted that the US would reach the peak number of cases by April 8, 2020 \cite{uhlig2020modeling}. Additionally, a sigmoidal Hill-type model predicts that the US will have 735,920 cases within 76 days of the outbreak, with 41,285 deaths \cite{aboelkassem2020Covid}. In contrast, our model predicts significantly more Covid-19 cases and deaths, with an extended duration past 2 years for the majority of states examined. Note: all disease models come with drawbacks. Our model is not mechanistic; it says nothing about Covid-19 specifically. While the quality of our fits to current data is very good, the accuracy of the projections depends on the reliability of the data used to generate the fit parameters.

Delays in transfers between compartments (such as our pseudo-quarantine), and in transferring between several compartments (effectively causing a delay) prior to re-entry in the susceptible population are known to cause oscillations in SIR type models \cite{hethcote1981nonlinear} -- reintroducing individuals into the susceptible compartment de-stabilizes the steady rate of infections. Temporary immunity, modeled by our reintroduction of recovered and excluded populations into the susceptible compartment, as well as relaxation of shelter-in-place orders, can produce yearly oscillations such as found in influenza and other human corona viruses (i.e. ``common colds'') \cite{kyrychko2005global, callow1990time}. It is quite likely that Covid-19 will become endemic in the United States with yearly spikes in cases.   

We have chosen to not include vaccination in our model, even though the race for development is currently underway \cite{le2020Covid}; rather we have focussed on non-medical interventions. Including such an effect is feasible in our model, where vaccinations would reduce the susceptible population. This could be at a constant rate, such as for newborn infants receiving the measles-mumps-rubella vaccine \cite{shulgin1998pulse}, or it could be a time dependent term reflecting people's confidence in vaccinations \cite{d2007bifurcation}. We look forward to seeing the effect of vaccination on our model predictions once Covid-19 vaccination has been demonstrated to be safe and effective in both animals and humans -- estimated to be possible by autumn 2021 \cite{peeples2020news}.  

In conclusion, we have developed a compartment model taking into account social distancing, undetected infecteds, and possible loss of immunity -- all issues which are relevant for Covid-19. The model describes current data very well for the states selected for study; this more realistic picture of the disease growth is likely due to both using a larger number of compartments than traditional SIR-type models, and to considering additional nonlinearity in the infectious power of the disease. While projections based on the model are not wholly optimistic, they do point to the fact that it is quite possible to avoid more severe outcomes with stronger measures -- increased social distancing, detection, and stay-at-home adherence -- than have been pursued so far.


%
\section*{CRediT authorship contribution statement}
{\bf Zeina Khan:} Conceptualization, Methodology, Writing, Visualization, Project administration. {\bf Frank Van Bussel:} Conceptualization, Methodology, Data Curation, Writing, Visualization, Software	Programming. {\bf Fazle Hussain:} Conceptualization, Methodology, Writing, Visualization, Supervision, Funding acquisition.

\section*{Declaration of Competing Interests}
The authors declare no competing interests.

\section*{Acknowledgements}
This study was supported by TTU President's Distinguished Chair Funds. We acknowledge early encouragement by Dr. Jamilur R. Choudhury to study Covid-19, and humbly dedicate this paper to his memory.

\newpage

\end{document}